\documentstyle[12pt,epsf]{article} 

\def\lsim{\mathrel{\lower2.5pt\vbox{\lineskip=0pt\baselineskip=0pt 
           \hbox{$<$}\hbox{$\sim$}}}} 
\def\gsim{\mathrel{\lower2.5pt\vbox{\lineskip=0pt\baselineskip=0pt 
           \hbox{$>$}\hbox{$\sim$}}}} 
\def\dis{\displaystyle}

\def\L{\Lambda}

\def\l{\lambda}

\begin{document} 
\begin{flushright}
DPNU-02-38\\ hep-th/0211268
\end{flushright}

\vspace{10mm}

\begin{center}
{\Large \bf 
 Correction terms to Newton law due to induced gravity in
 AdS background}

\vspace{20mm}
 Masato ITO 
 \footnote{E-mail address: mito@eken.phys.nagoya-u.ac.jp}
\end{center}

\begin{center}
{
\it 
{}Department of Physics, Nagoya University, Nagoya, 
JAPAN 464-8602 
}
\end{center}

\vspace{25mm}

\begin{abstract}
We calculate small correction terms to gravitational potential on
Randall-Sundrum brane with an induced Einstein term.
The behaviors of the correction terms depend on the magnitudes of $AdS$
radius $k^{-1}$ and a characteristic length scale $\l$ of model.
We represent the gravitational potential for arbitrary $k$ and $\l$ at all
distances.
\end{abstract} 

\newpage 

 Braneworld is based on the assumption that our four-dimensional world
 is embedded in higher-dimensional world.
 This framework shed light on an interpretation of the
 four-dimensional gravity, for instance,
 it is expected that weakness of the gravity we can feel should be
 explained by extra dimensions.
 In particular, it should be noted that localization of gravity occurs on
 a brane embedded in five dimensions.
 Recently two localized gravity models have been proposed. 
 In the Randall-Sundrum model \cite{Randall:1999ee,Randall:1999vf},
 extra dimension is non-compact, localization of gravity occurs on a
 flat $3$-brane embedded in five-dimensional anti-de Sitter space
 \cite{Lykken:1999nb,Karch:2000ct}.
 This is because zero mode of gravity becomes a bound state due to
 attractive force via positive tension brane.
 It is pointed out that the usual four-dimensional Newton law can be
 reproduced at distance which is larger than a radius of $AdS$ space.
 There have been several works with regard to localization of gravity
 on the brane in $AdS$ space
 \cite{Giddings:2000mu,Csaki:2000fc,Ito:2001nc,Ito:2002qp}.

 Model proposed by Dvali, Gabadadze and Porrati (DGP model) 
 \cite{Dvali:2000hr} consists of a $3$-brane embedded in
 five-dimensional Minkowski with infinite fifth dimension.
 Taking account of the induced four-dimensional Einstein term via
 quantum loop effects due to particles on the brane 
 \cite{Dvali:2000hr,Kofinas:2001qd,Kofinas:2001es},
 it is shown that gravitational potential becomes the usual Newton law 
 ($\sim 1/r$) at short distance and five-dimensional law 
 ($\sim 1/r^{2}$) at large distance.

 In \cite{Kiritsis:2002ca}
 it is shown that the effects of an induced Einstein term on
 a $3$-brane embedded in $AdS$ background play an important role in the
 modification of gravitational potential.
 Choosing appropriate value of radius of $AdS$ space,
 the five-dimensional gravity appears at intermediate distance and
 the four-dimensional gravity can be recovered at other distance.  

 The purpose of this letter is to investigate the behavior of gravity
 in above model.
 Although the contributions to Newton law of Kaluza-Klein modes
 are calculated in \cite{Kiritsis:2002ca}, in more detail,
 we calculate the exact form
 of correction terms to Newton law. 
 Moreover we estimate the correction terms for the region of $AdS$
 radius which is not studied in \cite{Kiritsis:2002ca}.
 Since the detection of deviation from Newton law 
 is performed by gravitational experiments at the present
 \cite{Hoyle:2000cv},
 it is important to estimate the correction terms.

 The low energy effective action we consider is given by  
\begin{eqnarray}
 S=\int d^{4}xdy\sqrt{-G}\left(\frac{M^{3}_{5}}{2}{\cal R}-\Lambda\right)
  +\int d^{4}x\sqrt{-g}\left(\frac{M^{2}_{4}}{2}R-V\right)
 \label{eqn1}\,,
 \end{eqnarray}
 where $M_{5}$ is the five-dimensional scale and $M_{4}$ is the 
 four-dimensional scale.
 Here $\L$ stands for negative bulk cosmological constant and $V$ is 
 brane tension.
 From (\ref{eqn1}) we can obtain the Einstein equation as follows
 \begin{eqnarray}
 {\cal R}_{MN}-\frac{1}{2}{\cal R}G_{MN}+\frac{\L}{M^{3}_{5}}G_{MN}
 +\l\delta(y)\left(R_{\mu\nu}-\frac{1}{2}Rg_{\mu\nu}
                    +\frac{V}{M^{2}_{4}}\right)
 \delta^{\mu}_{M}\delta^{\nu}_{N}=0
 \label{eqn2}\,,
 \end{eqnarray}
 where $\l=M^{2}_{4}/M^{3}_{5}$, we adopt indices as 
 $M,N=0,1,2,3,4$ and $\mu,\nu=0,1,2,3$.
 It is assumed that the fifth dimension $y$ is non-compact with
 $Z_{2}$ symmetry $y\sim-y$.
 The ansatz for metric is taken as follows
 \begin{eqnarray}
 ds^{2}=a^{2}(y)\eta_{\mu\nu}dx^{\mu}dx^{\nu}+dy^{2}
 \label{eqn3}\,,
 \end{eqnarray} 
 where $a(y)$ is warp factor.
 Solving the Einstein equation, we obtain warp factor and
 brane tension, namely, $a(y)=e^{-k|y|}$ and $V=6M^{3}_{5}k$, where
 $k=\sqrt{-\L/6M^{3}_{5}}$.
 
 In order to study the behavior of the gravitational potential on the
 brane, the gravitational fluctuations around the background metric are
 given by $\eta_{\mu\nu}+h_{\mu\nu}(x,y)$.
 Here, for simplicity, the tensor structure of gravity is neglected.
 Performing replacement of
 $h_{\mu\nu}(x,y)=h_{\mu\nu}(x)a^{-3/2}(y)\psi(z)$ imposed by
 $e^{k|y|}=1+k|z|$, the wave equation for fluctuation is given by 
 \begin{eqnarray}
 \left[-\frac{d^{2}}{dz^{2}}+\frac{15}{4\left(|z|+k^{-1}\right)^{2}}
       -\left(3k+\lambda m^{2}\right)\delta(z)
 \right]\psi(z)=m^{2}\psi(z)
 \label{eqn4}\,,
 \end{eqnarray}
 where $m^{2}$ is the four-dimensional mass which corresponds to
 the Kaluza-Klein mode.
 Note that $z$ coordinate becomes the conformally flat coordinate.
 The potential part of above equation is as same as one of the RS model,
 however, delta function part is explicitly different.
 This difference comes from localized kinetic term on the brane via
 induced gravity.
 From (\ref{eqn4}) the zero mode wave function $\psi_{0}(z)$ can be 
 normalizable, and we obtain $\psi_{0}(z)=k^{-1}(|z|+k^{-1})^{-3/2}$.
 Furthermore the wave function with KK-modes can be expressed in terms
 of linear combination of the Bessel functions as follows,
 \begin{eqnarray}
 \psi_{m}(z)=\sqrt{\frac{\dis m\left(|z|+\frac{1}{k}\right)}{1+A^{2}(m)}}
 \left\{
  Y_{2}\left(m\left(|z|+\frac{1}{k}\right)\right)
 +A(m)J_{2}\left(m\left(|z|+\frac{1}{k}\right)\right)
 \right\}\label{eqn5}\,,
 \end{eqnarray}
 where 
 \begin{eqnarray}
 A(m)=-\frac{\dis 2Y_{1}\left(\frac{m}{k}\right)
             +\lambda mY_{2}\left(\frac{m}{k}\right)}
 {\dis 2J_{1}\left(\frac{m}{k}\right)
  +\lambda mJ_{2}\left(\frac{m}{k}\right)}\label{eqn6}\,.
 \end{eqnarray}
 Here $A(m)$ is determined by the jump condition due to delta function
 at $y=0$ and the normalization factor is fixed by the
 orthonormalization condition of Bessel functions.

 Since we are interested in the correction terms to the four-dimensional
 Newton law between two unit masses on the brane,
 it is necessary to obtain the probability of gravity with KK-modes on
 the brane.
 From (\ref{eqn5}) we get
 \begin{eqnarray}
 \psi^{2}_{m}(0)&=&\frac{16}{\pi^{2}}\frac{k}{m}
 \left[4\left\{J^{2}_{1}\left(\frac{m}{k}\right)
 +Y^{2}_{1}\left(\frac{m}{k}\right)\right\}
 +\lambda^{2}m^{2}\left\{J^{2}_{2}\left(\frac{m}{k}\right)
 +Y^{2}_{2}\left(\frac{m}{k}\right)\right\}
 \right.\nonumber\\
 &&\hspace{2cm}\left.+4\lambda m
 \left\{J_{1}\left(\frac{m}{k}\right)J_{2}\left(\frac{m}{k}\right)
 +Y_{1}\left(\frac{m}{k}\right)Y_{2}\left(\frac{m}{k}\right)\right\}
 \right]^{-1}\label{eqn7}\,.
 \end{eqnarray}
 Here we used the Lommel's formula
 $J_{\nu+1}(x)Y_{\nu}(x)-J_{\nu}(x)Y_{\nu+1}(x)=2/\pi x$.
 The asymptotic behavior of $\psi^{2}_{m}(0)$ depends on the magnitude of
 argument in the Bessel functions, consequently,
 (\ref{eqn7}) is expressed as
 \begin{eqnarray}
 \psi^{2}_{m}(0)\sim
 \left\{\begin{array}{ll}
        \dis\frac{8}{\pi(4+\lambda^{2}m^{2})}&m\gg k\\
        &\\
        \dis\frac{m}{k(1+\lambda k)^{2}}&m\ll k\,.
        \end{array}\right. 
 \label{eqn8}\
 \end{eqnarray}
 Above equations can be derived by using the asymptotic form of Bessel
 functions, $J_{n}(x)\sim \sqrt{2/\pi x}\cos(x-(2n+1)\pi/4)$
 and $Y_{n}(x)\sim \sqrt{2/\pi x}\sin(x-(2n+1)\pi/4)$ for
 $x\gg 1$, $J_{n}(x)\sim x^{n}/2^{n}n!$ and 
 $Y_{n}(x)\sim -\pi^{-1}(2/x)^{n}$ for $x\ll 1$.
 The gravitational potential between two unit masses on the brane is
 expressed as
 \begin{eqnarray}
 V(r)=\frac{M^{-3}_{5}}{r}\psi^{2}_{0}(0)
      +\Delta V(r)\label{eqn9}\,,
 \end{eqnarray}
 where the first term is contribution of zero mode and the second term
 corresponds to the correction term which is generated by the exchange
 of KK-modes.
 Thus $\Delta V(r)$ is given by
 \begin{eqnarray}
 \Delta V(r)&=&
 M^{-3}_{5}\int^{\infty}_{0}dm\frac{e^{-mr}}{r}\psi^{2}_{m}(0)
 \label{eqn10}\,.
 \end{eqnarray}
 According to (\ref{eqn8}), it is necessary to divide the integral into 
 two regions, $m\ll k$ and $m\gg k$.
 Consequently we can get
 \begin{eqnarray}
 \Delta V(r)&\sim&
 M^{-3}_{5}\left\{
 \frac{1}{r}\int^{k}_{0}dm\frac{me^{-mr}}{k(1+\lambda k)^{2}}+
 \frac{1}{r}\int^{\infty}_{k}dm\frac{8e^{-mr}}{\pi(4+\lambda^{2}m^{2})}
 \right\}
 \label{eqn11}\,.
 \end{eqnarray}
 From (\ref{eqn9}) and (\ref{eqn11}), Newton law of zero mode plus
 KK-modes is calculated as follows
 \begin{eqnarray}
 V(r)&\sim& \frac{M^{-3}_{5}k}{r}\left(
 1+
 \frac{1}{k^{2}(1+\l k)^{2}}\frac{1-(1+kr)e^{-kr}}{r^{2}}
 \right.\nonumber\\
 &&\hspace{5cm}\left.
 +\frac{4}{\pi \l k}{\rm Re}
 \left[ie^{-i\frac{2r}{\l}}{\rm E_{i}}\left(i\frac{2r}{\l}-kr\right)
 \right]
 \right)\label{eqn12}\,,\nonumber\\
 \end{eqnarray}
 where ${\rm Ei}(-x)=-\int^{\infty}_{x}dt\;e^{-t}/t$
 is exponential integral function.
 Thus we can obtain the form of gravitational potential at
 arbitrary distance $r$.
 The asymptotic behavior of $V(r)$ depends on the magnitude of $\l k$.
 Below we consider the three cases $\l k\gg 1$, $\l k\ll 1$
 and $\l k\sim 1$, separately.

 For $\l k\gg 1$, we can get approximately the form of $V(r)$ at distance
 $r$  as follows,
 \begin{eqnarray}
 V(r)\sim
 \left\{
 \begin{array}{ll}
 \dis \frac{M^{-3}_{5}k}{r}
      \left(1+\frac{1}{(\l k^{2})^{2}r^{2}}\right)
 \sim\frac{M^{-3}_{5}k}{r} & r\gg k^{-1}\\ 
 & \\
 \dis\frac{M^{-3}_{5}k}{r}
      \left(1+\frac{4}{\pi \l k}
            \left(\gamma+\log(kr)\right)\sin\frac{2r}{\l}\right)
            \sim\frac{M^{-3}_{5}k}{r}& r\ll k^{-1}\,,
 \end{array}
 \right.\label{eqn13}
 \end{eqnarray}
 where $\gamma\sim 0.577$ is the Euler-Masceroni constant, and we used
 ${\rm E_{i}}(-x)\sim \gamma+\log x$ for $x\ll 1$.
 Note that the effective four-dimensional Planck scale can identified
 with $M^{2}_{p}\sim M^{3}_{5}/k$.
 Thus the leading correction term behaves as $r^{-2}$ at large
 distance and has logarithmic behavior at small distance.
 Since the correction terms are sufficiently suppressed,
 gravity behaves as four dimensions at whole range of $r$.

  For $\l k\ll 1$, from (\ref{eqn12}), we get 
  \begin{eqnarray}
 V(r)&\sim&
 \frac{M^{-3}_{5}k}{r}
 \left\{
 1+\frac{1-(1+kr)e^{-kr}}{k^{2}r^{2}}\right.\nonumber\\
 &&\left.\hspace{3cm}+\frac{4}{\pi \l k}
  \left[{\rm ci}\left(\frac{2r}{\l}\right)\sin\left(\frac{2r}{\l}\right)
       -{\rm si}\left(\frac{2r}{\l}\right)\cos\left(\frac{2r}{\l}\right)
 \right]
 \right\}\label{eqn14}\,,
 \end{eqnarray}
 Here we used the formula
 ${\rm Ei}(ix)={\rm ci}(x)+i{\rm si}(x)$, where
 ${\rm ci}(x)=-\int^{\infty}_{x}dt\;\cos t/t$ and
 ${\rm si}(x)=-\int^{\infty}_{x}dt\;\sin t/t$.
 At distance $r$ the forms of $V(r)$ are given by
 \begin{eqnarray}
 V(r)\sim
 \left\{
 \begin{array}{l}
 \dis \frac{M^{-3}_{5}k}{r}
      \left(1+\frac{1}{k^{2}r^{2}}+
      \frac{2}{\pi kr}\right)\sim\frac{M^{-3}_{5}k}{r} 
 \hspace{1cm} r\gg k^{-1}\\ 
  \\
 \dis \frac{M^{-3}_{5}k}{r}
      \left(1+\frac{2}{\pi kr}\right)\sim\frac{M^{-3}_{5}}{r^{2}}
 \hspace{2.5cm} \l\ll r \ll k^{-1}\\
  \\
 \dis\frac{M^{-3}_{5}k}{r}
      \left(1+\frac{2}{\l k}+
            \frac{4}{\pi \l k}\left[\gamma-1+\log\left(\frac{2r}{\l}\right)
            \right]\frac{2r}{\l}
            \right)\sim\frac{M^{-3}_{5}\l^{-1}}{r}\;\;\;\; r\ll \l\,.
 \end{array}
 \right.\label{eqn15}
 \end{eqnarray}
 Using ${\rm ci}(x)\sim \gamma+\log x\,,{\rm si}(x)\sim x-\pi/2$ for 
 $x\ll 1$ and ${\rm ci}(x)\sim \sin x/x\,,{\rm si}(x)\sim -\cos x/x$ for
 $x\gg 1$, above equations can be derived.
 In the case of intermediate distance $\l\ll r \ll k^{-1}$,
 since the contributions of KK-modes with $m\gg k$ can be dominant,
 gravity behaves as five dimensions.
 At distance $r\gg k^{-1}$ and $r\ll \l$,
 the four-dimensional Newton law can be reproduced.
 Moreover the effective four-dimensional Planck scale can be identified
 with $M^{2}_{p}\sim M^{3}_{5}/k$ for $r\gg k^{-1}$ and with
 $M^{2}_{p}\sim M^{3}_{5}\l=M^{2}_{4}$ for $r\ll \l$.
 Above results are consistent with \cite{Kiritsis:2002ca}.

 For a specific case, we choose $\l k=2$. 
 Consequently, $V(r)$ is taken the following form for $r\gg k^{-1}$ and
 $r\ll k^{-1}$,
 \begin{eqnarray}
 V(r)\sim
 \left\{
 \begin{array}{ll}
 \dis\frac{M^{-3}_{5}k}{r}\left(1+\frac{9}{k^{2}r^{2}}
 +\frac{e^{-kr}}{\pi kr}\right)& r\gg k^{-1}\\
 & \\
 \dis\frac{M^{-3}_{5}k}{r}\left(
 \frac{3}{2}
 +\frac{2}{\pi}\left(\gamma-1+\frac{1}{2}\log 2+
 \log kr\right)kr
 \right) & r\ll k^{-1}
 \end{array}
 \right.\,.\label{eqn16}
 \end{eqnarray}
 Thus gravity behaves as four dimensions at whole range of $r$.
 The behaviors of the correction terms at large distance are different from
 ones at small distance. 

 In summary we calculated the correction terms to Newton law
 on the $3$-brane with an induced Einstein term in $AdS$ background.
 The behavior of gravity depends on the magnitudes of $AdS$ radius
 $k^{-1}$, a characteristic length scale $\l$ and distance $r$.
 Remarkably, for $\l k\ll 1$ the five-dimensional gravity appears at
 intermediate scale $\l\ll r\ll k^{-1}$.
 For $\l k\gg 1$ and $\l k\sim 1$ gravity behaves as four dimensions
 at whole distance, however, the behaviors of correction terms
 at large distance are different from ones at small distance.

 Since the possibility of theories with extra dimensions is indicated,
 the experiments of searching for the presence of extra dimensions are 
 increasingly performed.
 From recent gravitational experiments, it is found that the
 gravitational force $r^{-2}$ law is maintained up to $0.218$mm
 \cite{Hoyle:2000cv}.
 However, it is unknown whether $r^{-2}$ law is violated or not at
 micrometer range.
 In near future, it is expected that the sophisticated equipment
 of gravitational experiment will confirm small correction terms
 shown in (\ref{eqn13}), (\ref{eqn15}) and (\ref{eqn16}).

%%%%%%%%%%%%%%%%%%%%%%%%%%%% reference %%%%%%%%%%%%%%%%%%%%%%%%%
%


\begin{thebibliography}{99}
%
%
 %\cite{Randall:1999ee}
 \bibitem{Randall:1999ee}
 L.~Randall and R.~Sundrum,
 ``A large mass hierarchy from a small extra dimension,''
 Phys.\ Rev.\ Lett.\  {\bf 83}, 3370 (1999) [hep-ph/9905221].
%
 %\cite{Randall:1999vf}
 \bibitem{Randall:1999vf}
 L.~Randall and R.~Sundrum,
 ``An alternative to compactification,''
 Phys.\ Rev.\ Lett.\  {\bf 83}, 4690 (1999) [hep-th/9906064].
%
 %\cite{Lykken:1999nb}
 \bibitem{Lykken:1999nb}
 J.~Lykken and L.~Randall,
 ``The shape of gravity,'' JHEP {\bf 0006}, 014 (2000) [hep-th/9908076].
%
 %\cite{Karch:2000ct}
 \bibitem{Karch:2000ct}
 A.~Karch and L.~Randall,
 ``Locally localized gravity,'' JHEP {\bf 0105}, 008 (2001)
 [hep-th/0011156].
%
%
 %\cite{Giddings:2000mu}
 \bibitem{Giddings:2000mu}
 S.~B.~Giddings, E.~Katz and L.~Randall,
 ``Linearized gravity in brane backgrounds,'' JHEP {\bf 0003}, 023 (2000)
 [hep-th/0002091].
%
 %\cite{Csaki:2000fc}
 \bibitem{Csaki:2000fc}
 C.~Csaki, J.~Erlich, T.~J.~Hollowood and Y.~Shirman,
 ``Universal aspects of gravity localized on thick branes,''
 Nucl.\ Phys.\ B {\bf 581}, 309 (2000) [hep-th/0001033].
%
 %\cite{Ito:2001nc}
 \bibitem{Ito:2001nc}
 M.~Ito,
 ``Newton's law in braneworlds with an infinite extra dimension,''
 Phys.\ Lett.\ B {\bf 528}, 269 (2002) [hep-th/0112224].
%
 %\cite{Ito:2002qp}
 \bibitem{Ito:2002qp}
 M.~Ito,
 ``Localized gravity on de Sitter brane in five dimensions,''
 [hep-th/0204113].
%
 %\cite{Dvali:2000hr}
 \bibitem{Dvali:2000hr}
 G.~R.~Dvali, G.~Gabadadze and M.~Porrati,
 ``4D gravity on a brane in 5D Minkowski space,''
 Phys.\ Lett.\ B {\bf 485}, 208 (2000)
 [arXiv:hep-th/0005016].
%
 %\cite{Kofinas:2001qd}
 \bibitem{Kofinas:2001qd}
 G.~Kofinas, E.~Papantonopoulos and I.~Pappa,
 ``Spherically symmetric braneworld solutions with (4)R term in the bulk,''
 arXiv:hep-th/0112019.
%
 %\cite{Kofinas:2001es}
 \bibitem{Kofinas:2001es}
 G.~Kofinas,
 ``General brane cosmology with (4)R term in (A)dS(5) or Minkowski bulk,''
 JHEP {\bf 0108}, 034 (2001)
 [arXiv:hep-th/0108013].
%
 %\cite{Kiritsis:2002ca}
 \bibitem{Kiritsis:2002ca}
 E.~Kiritsis, N.~Tetradis and T.~N.~Tomaras,
 ``Induced gravity on RS branes,''
 JHEP {\bf 0203}, 019 (2002)
 [arXiv:hep-th/0202037].
%
 %\cite{Hoyle:2000cv}
 \bibitem{Hoyle:2000cv}
 C.~D.~Hoyle, U.~Schmidt, B.~R.~Heckel, E.~G.~Adelberger,
 J.~H.~Gundlach, D.~J.~Kapner and H.~E.~Swanson,
 ``Sub-millimeter tests of the gravitational inverse-square law: 
 A search  for 'large' extra dimensions,''
 Phys.\ Rev.\ Lett.\  {\bf 86}, 1418 (2001) [hep-ph/0011014].
%
\end{thebibliography}
\end{document}